
\font\rmu=cmr10 scaled\magstephalf
\font\bfu=cmbx10 scaled\magstephalf

\font\it=cmti10 scaled \magstephalf
\font\bf=cmbx10 scaled\magstephalf
\rmu

\font\rmus=cmr8
\font\rmuss=cmr6
\font\mait=cmmi10 scaled\magstephalf
\font\maits=cmmi7 scaled\magstephalf
\font\maitss=cmmi7
\font\msyb=cmsy10 scaled\magstephalf
\font\msybs=cmsy8 scaled\magstephalf
\font\msybss=cmsy7
\font\bfus=cmbx7 scaled\magstephalf
\font\bfuss=cmbx7
\font\cmeq=cmex10 scaled\magstephalf

\textfont0=\rmu
\scriptfont0=\rmus
\scriptscriptfont0=\rmuss

\textfont1=\mait
\scriptfont1=\maits
\scriptscriptfont1=\maitss

\textfont2=\msyb
\scriptfont2=\msybs
\scriptscriptfont2=\msybss

\textfont3=\cmeq
\scriptfont3=\cmeq
\scriptscriptfont3=\cmeq

\newfam\bmufam  \textfont\bmufam=\bfu
      \scriptfont\bmufam=\bfus \scriptscriptfont\bmufam=\bfuss

\hsize=15.5cm
\vsize=21cm
\baselineskip=16pt   
\parskip=12pt plus  2pt minus 2pt

\def\semi{\bigcirc\kern-1em{s}\;}

\def\R{{\rm I\!R}}

\def\Q{{\mathchoice
{\setbox0=\hbox{$\displaystyle\rm Q$}\hbox{\raise 0.15\ht0\hbox to0pt
{\kern0.4\wd0\vrule height0.8\ht0\hss}\box0}}
{\setbox0=\hbox{$\textstyle\rm Q$}\hbox{\raise 0.15\ht0\hbox to0pt
{\kern0.4\wd0\vrule height0.8\ht0\hss}\box0}}
{\setbox0=\hbox{$\scriptstyle\rm Q$}\hbox{\raise 0.15\ht0\hbox to0pt
{\kern0.4\wd0\vrule height0.7\ht0\hss}\box0}}
{\setbox0=\hbox{$\scriptscriptstyle\rm Q$}\hbox{\raise 0.15\ht0\hbox to0pt
{\kern0.4\wd0\vrule height0.7\ht0\hss}\box0}}}}
\def\C{{\mathchoice
{\setbox0=\hbox{$\displaystyle\rm C$}\hbox{\hbox to0pt
{\kern0.4\wd0\vrule height0.9\ht0\hss}\box0}}
{\setbox0=\hbox{$\textstyle\rm C$}\hbox{\hbox to0pt
{\kern0.4\wd0\vrule height0.9\ht0\hss}\box0}}
{\setbox0=\hbox{$\scriptstyle\rm C$}\hbox{\hbox to0pt
{\kern0.4\wd0\vrule height0.9\ht0\hss}\box0}}
{\setbox0=\hbox{$\scriptscriptstyle\rm C$}\hbox{\hbox to0pt
{\kern0.4\wd0\vrule height0.9\ht0\hss}\box0}}}}

\font\fivesans=cmss10 at 4.61pt
\font\sevensans=cmss10 at 6.81pt
\font\tensans=cmss10
\newfam\sansfam
\textfont\sansfam=\tensans\scriptfont\sansfam=\sevensans\scriptscriptfont
\sansfam=\fivesans
\def\sans{\fam\sansfam\tensans}
\def\Z{{\mathchoice
{\hbox{$\sans\textstyle Z\kern-0.4em Z$}}
{\hbox{$\sans\textstyle Z\kern-0.4em Z$}}
{\hbox{$\sans\scriptstyle Z\kern-0.3em Z$}}
{\hbox{$\sans\scriptscriptstyle Z\kern-0.2em Z$}}}}

\def\Journal#1#2#3#4{{#1} {\bf #2} (#4) #3}

\def\NCA{\it Nuovo Cimento}
\def\NPB{{\it Nucl. Phys.} B}
\def\PLB{{\it Phys. Lett.}  B}
\def\PRL{\it Phys. Rev. Lett.}
\def\PRD{{\it Phys. Rev.} D}
\def\CQG{\it Class. Quant. Grav.}
\def\MPLA{{\it Mod. Phys. Lett.} A}
\def\CMP{\it Comm. Math. Phys.}
\def\JMP{\it J. Math. Phys.}
\def\JGP{\it J. Geom. Phys.}
\def\JFA{\it J. Funct. Anal.}
\def\AM{\it Adv. Math.}

\newcount\foot
\foot=1
\def\note#1{\footnote{${}^{{}^{\number\foot}}$}{\ftn #1}\advance\foot by 1}

\def\frac#1#2{{#1\over #2}}
\def\text#1{\quad{\hbox{#1}}\quad}

\font\ch=cmbx12 scaled\magstephalf
\font\ftn=cmr8 scaled\magstephalf

\font\it=cmti10 scaled\magstephalf
\font\bf=cmbx10 scaled\magstephalf
\font\titch=cmbx12 scaled\magstep2
\font\titname=cmr10 scaled\magstep2
\font\titit=cmti10 scaled\magstep1
\font\titbf=cmbx10 scaled\magstep2

\nopagenumbers

\line{\hfil AEI-025}
\line{\hfil January 14, 1997}
\vskip4cm
\centerline{\titch Still on the way to quantizing gravity \note{Invited 
talk at the 12th 
Italian Conference on General Relativity and Gravitational Physics,
Roma, September 23-27, 1996.}}
\vskip3.5cm
\centerline{\titname R. Loll}
\vskip.5cm
\centerline{\titit Max-Planck-Institut f\"ur Gravitationsphysik}
\vskip.2cm
\centerline{\titit Schlaatzweg 1}
\vskip.2cm
\centerline{\titit D-14473 Potsdam, Germany}

\vskip3.5cm
\centerline{\titbf Abstract}
\vskip0.2cm

I review and discuss some recent developments in non-perturbative 
approaches to quantum gravity, with an emphasis on discrete formulations, 
and those coming from a classical connection description.

\vfill\eject
\footline={\hss\tenrm\folio\hss}
\pageno=1

\line{\ch 1 Quantizing gravity\hfil}

The subject of this talk are several more or less developed research
programs for constructing a consistent quantum theory of Einstein gravity
in four space-time dimensions. I will focus on quantizations of pure
gravity, as opposed to unified approaches like string theory.
This assumes that one can extract physical information
by quantizing gravity first, and adding matter fields in a second
step, a point of view that is maybe not universally shared.
I will concentrate on so-called non-perturbative approaches,  
which avoid the decomposition of the space-time metric 
$g_{\mu\nu}$ into a flat Minkowskian piece $\eta$ plus small 
fluctuations, $g_{\mu\nu}=\eta_{\mu\nu}+\kappa h_{\mu\nu}$. 
However, note there are claims that in spite of its non-renormalizability, 
gravity may be treated perturbatively as a quantum effective field 
theory, valid only in a limited energy range [1].

One alternative to a perturbative approach is to 
describe gravity in a way involving from the outset contributions from 
arbitrarily curved space-times, and not only from those close to the 
flat Minkowski metric. Since the space of all (pseudo-Riemannian)
geometries has a complicated structure, one may in a first step
try to simplify the problem by approximating it by a discrete space. 
In addition, the underlying space-time itself may be discretized. 
One fundamental difficulty
with this kind of ansatz is how to implement (in an approximate
sense) the diffeomorphism symmetry of the classical theory.

Some of the discrete approaches to quantum gravity
have been around for many years, while others are of a more recent origin. 
I will focus on some specific developments that have their root 
in a new Hamiltonian reformulation of continuum gravity, which is based on 
connection variables, although I will mention some results from alternative 
research programs in passing. 

The use of connection variables in gravity, especially in
first-order formulations, is not new (see [2] for some recent applications).
Often the aim of such reformulations is to make
gravity resemble a gauge theory as closely as possible. In this sense,
the arguably most successful classical proposal, or at least the one 
that has had a big impact on the construction of a quantum theory, is  
due to Ashtekar [3], 
and is based on $sl(2,\C)$-valued connection forms. There
also exists a real version of his variables [4], 
whose virtues will be described later. 
They are $su(2)$-valued, and the gauge algebra is therefore truly minimal.
Since the classical gravitational phase space (before considering
the dynamics and diffeomorphism symmetry) is in this formulation identical 
with that of a Yang-Mills theory, it is suggestive to attempt
a discretization along the lines of Hamiltonian lattice gauge theory [5].
Progress on this will be reported below.
Indeed, many constructions of the non-perturbative {\it continuum} quantum
theory closely resemble those of a lattice theory, and consequently have led
to the appearance of certain discrete features also in this case.

\vskip2cm
\line{\ch 2 Discrete approaches to quantum gravity\hfil}

Considering our incomplete understanding of quantum gravity, it is 
probably most fruitful to consider the approaches discussed below
as complementary. 

\vskip1cm
\line{\bf 2.1 Quantum Regge calculus\hfil}

The classical starting point for this quantization method is the
approximation of a metric space-time by a simplicial manifold
with consistent edge length assignments $l_i$, carrying the 
information about components of the metric tensor.
The Einstein-Hilbert action is written, following Regge [6],
as a functional of the edge lengths only. 

The quantization proceeds via a path
integral over all allowed edge length configurations for a given
simplicial manifold [7]. In order to have the (Euclidean) action 
bounded from below, one usually adds higher-order curvature terms to the
action when performing numerical simulations. There is an ongoing
discussion about the correct path-integral measure, and whether or not
it should include a (possibly non-local) Faddeev-Popov determinant 
(see, for example, [8] and references 
therein). Recall that the form of the measure 
should encode the fact that one is integrating over a discrete 
analogue of the space of Riemannian structures modulo gauge, i.e.
the space ${\rm Riem}M/{\rm Diff} M$. As far as I am aware, there is
as yet no analytical derivation of a Regge calculus measure in four 
dimensions. In practice, a few simple
measures are in use, such as $\prod_i d(l_i)^2$. In Monte-Carlo
simulations one finds evidence for a non-trivial phase 
structure [9]. Because
of the complexity of the calculations, not a great deal of data is available
yet.

\vskip1cm
\line{\bf 2.2 Dynamical triangulations\hfil}

In this variant of the Regge calculus program, one again uses simplicial
mani\-folds and the Regge form of the action, but a different space of quantum
configurations in the path integral. Instead of summing over all possible
edge lengths for a fixed, simplicial manifold, one fixes a fiducial length
for all 1-simplices, and sums over all distinct triangulations of 
$M$ [10].
Information about the geometry is now encoded combinatorially: the
fact that ``length" comes only in discrete bits implies that some derived
quantities, like curvatures and volumes, can be obtained by simply counting
simplices of a certain dimension, which is extremely convenient
from a numerical point of view. By varying triangulations in numerical 
simulations, one tries to sample uniformly the space of (discrete) metric
structures.  

A recent analytical construction (using spaces of bounded Riemannian 
geometries) provides a theoretical understanding of some properties
of the dynamical triangulations approach, for example, the 
diffeomorphism-invariance of its path-integral measure.
This enables one to evaluate the partition function
asymptotically for large triangulations, and leads to a prediction for
the location of the critical point, corresponding to a higher-order phase 
transition [11]. The analytical predictions are in good agreement with
Monte-Carlo simulations, wherever available.
Comparing to the situation found in other statistical systems,
and considering the fact that we are in four dimensions, this seems almost
too good to be true.
 
However, one has to remember that the discrete path integral approaches
described so far deal with the Euclidean theory. It is not at all obvious
how the results could be ``analytically continued" to the correct
signature regime, given that this sector is basically inaccessible
numerically. 

\vskip1cm
\line{\bf 2.3 Ashtekar gravity on the lattice\hfil}

The signature problem is avoided by going to a Hamiltonian formulation. 
Unfortunately, the (3+1)-projected form of the algebra of 4-dimensional 
space-time
diffeomorphisms loses much of its simplicity, which also causes problems
in discretized approaches. In addition, as far as numerical simulations
are concerned, there is no Hamiltonian method matching the
efficiency of the Monte-Carlo methods of the path integral approach.
(This is why nobody performs QCD computer simulations in terms of Hilbert
spaces and operator algebras.) The application to generally covariant
theories is largely uncharted territory. 

If one wants to use Ashtekar-type variables {\it and} 
exploit the resemblance with lattice gauge theory,
there is as yet no alternative to the Hamiltonian formulation.
The reason for this is that the 
Einstein-Hilbert action, unlike the Yang-Mills action,
cannot be written purely as a functional of a four-dimensional gauge potential.
The close kinematical similarity between connection gravity and gauge 
theory holds only at the canonical level. 
A natural discretization of the Hamiltonians formulation consists in
the approximation of spatial slices by three-dimensional lattices, with ``time" 
left continuous.
(A somewhat different point of view is taken by Reisenberger, who studies 
Euclidean quantum gravity as a discretized path integral over configurations 
of two-surfaces in space-time [12].) 

The dynamics of Yang-Mills theory and gravity on the lattice are of course
totally diffe\-rent. The role of the quantum Hamiltonian is played in gravity
by the Hamiltonian constraint. Consequently, the problem of
diagonalizing the Hamiltonian is replaced by that of finding the
states that are annihilated by the quantum Hamiltonian operator, i.e.
the eigenvectors with eigenvalue zero. In addition, as mentioned earlier,
in lattice gravity one has to incorporate some ``remnant" of the 
diffeomorphism symmetry. There are at least two ways of doing this: either 
using discrete versions of the diffeomorphism constraints to project out
physical lattice states, in which case one should check that the quantum
constraint algebra is free of anomalies in the continuum limit. 
Alternatively, one may try to define a lattice measure that goes over to a 
diffeomorphism-invariant measure in the continuum limit. 
These important issues are momentarily under study. What has already been
explored on the lattice are the spectra of certain geometric 
operators [13-16]. 
Also a self-adjoint Hamiltonian 
operator can be defined rigorously, in spite of its 
non-polynomiality [17], but nothing so 
far is known about its spectrum. 

\vskip1cm
\line{\bf 2.4 Ashtekar gravity in the continuum\hfil}

The reason why I mention this approach here together with the discrete
ones, is the pre\-sence of certain discrete features mentioned earlier.
They enter because of the unconventional way in which one defines the
quantum theory, which is based on an algebra of one-dimensional objects, 
so-called Wilson loops. Related to this, there has been a cross-fertilization
of ideas with the lattice formulation, although the overall outlook of
the two is rather different. In a further development of ideas that first
arose within the so-called loop representation, it turns out that a
convenient way of labelling quantum states in the continuum theory
includes all possible imbedded lattices or graphs (i.e. objects obtained
by gluing one-dimensional imbedded edges). For example,
a quantum state can be given by fixing a lattice and specifying a few
further data for its edges and vertices. The entire Hilbert space therefore 
looks like a tensor product of all possible lattice theories. 
Note, however, that these lattices are all
imbedded in the smooth 3-manifold $\Sigma$, and one can therefore  
define a natural action of ${\rm Diff}\Sigma$ on them. By contrast, the
lattice of the discrete lattice approach is not imbedded anywhere, but
rather itself an approximation to $\Sigma$. 

The discreteness alluded to above refers to the discreteness of the 
spectra of certain geometric operators one can define in the continuum theory, 
associated with spatial volumes, areas and 
lengths [18,19]. This 
comes about because the action of these operators on typical loop or
graph states is partly combinatorial, in the sense that it reduces to
discrete (and typically finite) sums over points of the graph, and leads
to rearrangements of edges at these intersection points. The discreteness
of the geometric spectra is often taken to imply  
that geometry is quantized, in the same way as spin is quantized
in quantum electro-dynamics. This is somewhat at odds with the fact that
at various points of the construction of the continuum quantum theory one uses 
the smooth structure of the background manifold $\Sigma$ in a crucial way.

An interesting feature of this quantum representation is that not only 
geometric operators can be made well-defined and are
finite after some appropriate three-dimensional smearing, with no
renormalization necessary, but there also exist Hamiltonian operators 
with the same properties [20]. 
This is sometimes taken as an indication 
that the quantum theory is complete as it stands, without any further need for
a continuum limit, a point of view that has been 
criticized by Smolin [21].

\vskip2cm
\line{\ch 3 Brief outline of the connection approach\hfil}

Usually, one thinks of gravity as describing space-time geometries, 
and expresses solutions to the Einstein equations in terms of their
line elements $ds^{2}=g_{\mu\nu}dx^{\mu}dx^{\nu}$, suggesting the use 
of the metric tensor $g_{\mu\nu}(x)$ as a basic field-theoretic 
quantity. Since quantization attempts using the  $\hat g_{\mu\nu}(x)$ 
as basic operators have run into serious difficulties, one might 
wonder whether there is some more fundamental set of classical variables 
that could form a more suitable starting point for the quantization. 
This may be compared to the case of electro-magnetism, where the 
requirements of locality and Lorentz covariance suggested the use of the
gauge potentials $A_{\mu}$ as basic variables in the perturbative 
quantization, in place of the gauge-invariant components of the field
strength tensor, which however have a more direct physical 
interpretation classically. 

For the case of gravity, there exists a well-known paradigm in
three space-time dimensions, where one may reformulate Einstein 
gravity as a Chern-Simons [22] 
or gauge theory [23], with actions of the form
$S[A]=\int A\wedge dA +A\wedge A\wedge A$ and $S[A,e]=\int e\wedge 
F(A)$ respectively, where the $A$'s are different connection-forms, with 
curvature $F$, and $e$ is a dreibein variable. 
This has led to a new explicit representation of the physical, reduced phase 
space, and to new insights into its quantization. 

The reason why one may gain anything by a simple change of classical
variables is partly due to the fact that gravity is a constrained 
theory, i.e. its initial data in terms of any set of basic variables 
cannot be specified freely, but is subject to a number of constraints.
Different ways of writing these constraints (although classically equivalent) 
can in principle lead to inequivalent quantum theories 
or at least suggest different ways of setting up the quantization. 

The idea to rewrite four-dimensional Einstein gravity in the form

$$
S^{\rm Einst}[A,e]=\int d^{4}x\, e\wedge e\wedge F(A),\eqno(1)
$$

\noindent with a selfdual $SO(4,\C)$-connection $A_{\mu}^{I}$ and a 
vierbein $e_{\mu}^{I}$ is due to Ashtekar [3], and predates the
developments in the three-dimensional theory mentioned earlier. The 
true benefits of this formulation become apparent after the
3+1 decomposition has been performed. In a nutshell, these advantages
are

\item{(A1)} all constraint equations are of the form of low- (up
to fourth) order polynomials on phase space, implying a potentially vast 
simplification in the operator quantization of these expressions;

\item{(A2)} the canonical variable pairs $(A_{a}^{i},E^{a}_{i})$ are
Yang-Mills variables, taking values in the gauge algebra 
$so(3,\C)\equiv sl(2,\C)$, subject to the Gauss law constraints 
${\cal D}_{a}(A)E^{a}_{i}=0$.
This enables one to import quantization techniques from gauge field
theory.

\noindent Unfortunately, these very useful properties come at a price, since

\item{(D1)} $A_{a}^{i}$ takes values in a {\it complex} algebra; 
in order to recover real gravity, 
a set of reality conditions has to be imposed.

The complicated functional form of these reality conditions makes it
very hard to implement them in the quantum theory, and this problem 
has so far remained unsolved. To put it simply: (D1) spoils (A1) and
(A2)! Fortunately, this does not mean that the entire approach is doomed, as
there is a closely related real connection formulation available, 
which avoids the disadvantage (D1), while retaining (A2) and part of (A1). 

\vskip1cm
\line{\bf 3.1 The new trend towards reality\hfil}

The credit for promoting alternative real formulations of gravity, and
in particular the real connection approach described below should go to
Barbero [24,4].
A unified derivation for both the complex and the real
Ashtekar variables starts from the canonical $SO(3)$-ADM variables 
$(E_i^a,K^i_a)$, with $E_i^a$ a densitized inverse dreibein,
$E_i^a=\sqrt{\det e}\, e_i^a$, where $e_a^i e_{bi}=g_{ab}$, 
$e_{a}^{i}e^{b}_{i}=\delta_{a}^{b}$, and $K^i_a$ is
the extrinsic curvature with one spatial index converted, 
$K^i_a=K_{ab}\, e^{bi}$. There are two ways of obtaining a Yang-Mills
variable pair $(A_{a}^{i},E^{a}_{i})$ by a canonical transformation on 
$(E_i^a,K^i_a)$. Choosing $i{\cal F}$ as a symplectic infinitesimal generator 
leads to the complex Ashtekar variables $A^{\C}=\Gamma+iK$, whereas the purely 
real generator ${\cal F}$ leads to a real connection variable 
$A^{\R}=\Gamma+K$. ${\cal F}$ is 
the spatial integral ${\cal F}[E]=\int d^3x\, \Gamma_a^i E^a_i$, where the
spin connection $\Gamma$ is a complicated function of the $E^a_i$, defined by
${\cal D}_a(\Gamma)E^b_i=0$.  

One may introduce a one-parameter family of canonical transformations [4], 
labelled by a complex parameter $\beta\not=0$, 
corresponding to connections ${A^{(\beta)}}_a^i=\Gamma_a^i+
\beta K_a^i$, with Poisson brackets 
$\{ A^{(\beta)}{}_a^i(x),E^b_j(y)\}=-\beta\delta^i_j\delta_a^b\delta^3(x,y)$.
Note that only Barbero's choice,  $\beta=-1$, leads to a genuine
canonical transformation. One can construct a corresponding action principle
with a free parameter, giving rise to these different Hamiltonian
formulations [25]. The reality conditions necessary in the complex 
approach (with $\beta=\pm i$) are given by
$A^{\C}+A^{\C*}=2\Gamma (E)$, and are non-polynomial in $E$ 
(since $\Gamma$ is). Alternative polynomial forms exist, but have not
turned out useful in the quantization. 

After the $\beta$-dependent canonical transformation, the
classical (Gauss law, diffeomorphism and Hamiltonian) constraint
equations take the form

$$
\eqalignno{
G_i&:=\nabla^{(\beta)}_aE^a_i=0,&(2)\cr
V_a&:=F^{(\beta)}{}_{ab}^i E^b_i=0,&(3)\cr
H&:=\epsilon^{ijk}E^a_i E^b_j\, (F^{(\beta)}_{abk}-(\frac{1}{\beta^2}+1)
\, R_{abk})=0,&(4)}
$$

\noindent where $\nabla^{(\beta)}$ denotes 
the covariant derivative with respect
to the connection $A^{(\beta)}$, and $F^{(\beta)}$ and $R$ are the field 
strengths of the connections $A^{(\beta)}$ and $\Gamma$,

$$
\eqalignno{
F^{(\beta)}{}_{ab}^i&=2 \partial_{[a}A^{(\beta)}{}_{b]}^i +\epsilon^{ijk}
 A^{(\beta)}_{aj} A^{(\beta)}_{bk},&(5)\cr
R_{ab}^i&=2 \partial_{[a}\Gamma_{b]}^i +\epsilon^{ijk} \Gamma_{aj}
\Gamma_{bk}.&(6)}
$$

\noindent Note that the only explicit $\beta$-dependence occurs in
the second term of the Hamiltonian constraint. The simplification
associated with Ashtekar's choice $\beta=\pm i$ is the
vanishing of the non-polynomial, $\Gamma$-dependent curvature term,
leading to a polynomial form for all of
the constraints. For all other choices of $\beta$, the Hamiltonian retains 
a complicated potential
term. Since the functional form of the other constraints does not change
with $\beta$, their quantization is unaffected, at least formally
(in reality, as already pointed out, a complex value for
$\beta$ leads to complications in defining the quantum theory; 
many structures existing for real gauge theories cannot easily be
translated to complex ones).

Can the complicated Hamiltonian of the real connection be quantized in
some way, or are we back to the problems besetting the old
ADM-style quantizations? The surprising answer is that one seems to
be able to do better in the connection approach. This has to do with
some features of the quantum theory, and ultimately with the fact that
the metric configuration variables (or, equivalently, the inverse
dreibeins) become differential operators in Yang-Mills-inspired
representations for quantum gravity, and not multiplication 
operators (as happens for the components of the metric tensor in other
representations).

Additionally, in a loop representation, based on quantum analogues of
one-dimensional Wilson loops, the operator corresponding to the
(local or integrated) squared volume function 
$\det E\equiv\det g$, is a relatively simple polynomial, self-adjoint 
operator, whose spectrum can be computed explicitly.
This allows one in principle to quantize arbitrary functions of $\det g$ in 
terms of a volume eigenbasis, a property that has no analogue in the old
metric representations. Exploiting this idea, a discretized version of
$H$ can be rigorously quantized on the lattice [17]; the same 
can be done for the rescaled Hamiltonian $(\det E)^{-1/2}H$ in the
continuum [20].

\vskip1cm
\line{\bf 3.2 Loop quantum gravity\hfil}

The idea of basing a non-perturbative formulation of quantum gravity
on non-local gauge-invariant 
Wilson loop variables goes back to Rovelli and Smolin [26].
Classical Wilson loops depend on a connection form $A$ and a closed
curve $\gamma$ in $\Sigma$, and are defined as the trace of the 
path-ordered integral of $A$ (a ``parallel transport" or ``holonomy" matrix
$U_\gamma$) along $\gamma$, 
${\rm Tr}\,U_\gamma[A]\equiv{\rm Tr\, P}\exp\oint_\gamma A$.
The classical (commutative) algebra of the phase space functions 
${\rm Tr}\,U_\gamma$ is then promoted to an algebra of operators
$\hat {\rm Tr}\,U_\gamma$. In addition, one needs to quantize phase space
functions depending on the fields $E$. The notion of what constitutes a
suitable set of momenta in this approach has since then 
undergone several modifications.

Ashtekar and Isham initiated the program of making 
the kinematical framework rigorous, in a $C^*$-algebra 
approach [27]. (This only
works for a real, not a complex gauge algebra, but at the time 
real connections for Lorentzian gravity were not yet being considered.)
This was followed by a series of papers, characterizing the quantum
configuration space (which is labelled by sets $(\gamma_1,\dots,\gamma_n)$
of imbedded loops), and defining measures,
integration, differential operators and other natural structures on this 
space [28], again for the case of real connections.
As mentioned in the beginning, many of these constructions are reminiscent
of lattice gauge theory. One must however remember
that the requirement of having well-defined Wilson loop operators 
is a non-trivial assumption in the continuum theory, and singles out
a particular kind of quantum representation. One may argue that it is
more physical to base the quantization on a set of truly three-dimensional
objects, such as flux tubes [29].

Who has followed the developments in this area of research
may wonder why in many discussions the ``loop representations" have been 
substituted by ``spin networks representations". 
This is only related to a shift in emphasis. In the early days,
``loops" where mostly associated with non-intersecting loops (but possibly
non-trivially linked or knotted), 
partly due to the fact that the first solutions
to the Wheeler-DeWitt equation in the loop representation were labelled
by such loops. Later there was a growing realization that intersections
are crucial, indeed, it was shown that quantum states with non-zero
volume eigenvalues must necessarily be labelled by loop configurations
containing intersections of valence at least four [13]. 
With the new emphasis on intersecting Wilson loops, a new way of labelling 
quantum states, by so-called
spin-network states [30], became convenient. 
They are given by particular anti-symmetrized linear combinations
of Wilson loop states sharing the same support. 
 
\vskip1cm
\line{\bf 3.3 Geometric operators\hfil}

A set of operators that can be conveniently studied in terms of a spin-network
basis are the geometric operators associated
with the classical volume, area and length functions. 
They may seem of no immediate physical relevance for the full
quantum theory, because they commute neither with the Hamiltonian nor
the diffeomorphism constraints. However, at least the volume operator turns
out to be useful in defining quantum operators depending on non-polynomial
phase space functions, that classically can be written as
polynomials up to arbitrary powers of the determinant of the metric,
like, for instance, the Hamiltonian constraint of the real connection approach.

Geometric operators can be regularized and are finite
with discrete spectra in the loop 
representation [18,19] (see [15] for a more complete bibliography).
Again this result is intimately tied to the fact that wave
functions carry discrete loop labels. The action of a geometric operator
on such a loop state can be replaced by a discrete sum over intersection
points of the underlying loop, and the operator action reduced to
a single intersection is purely quantum-mechanical.

Discretized analogues of the geometric operators can be defined in the
lattice theory [13-16]. 
It turns out that certain results obtained 
on the lattice are relevant to the continuum theory, 
because the operator actions
can be identical whenever one evaluates a continuum operator on a
{\it fixed} loop or graph configuration which can be 
realized as a subset of edges of a cubic (imbedded) lattice. 
Thus one may compare the operator spectra, at least partially. 

During lattice investigations it was found that
geometric operators do not in general commute [15]. 
This is surprising because the classical geometric functions 
depend only on half of the canonical variables, the inverse dreibeins
$E^a_i$, and therefore Poisson-commute. The 
non-commutativity on the lattice is a result of the choice of basic variables,
namely, as non-local versions (integrated over edges) of the pair $(A,E)$.
Not even the smeared-out versions of the $E$'s commute among themselves,
because ``integrating $E$ along a link" involves parallel transport and
therefore knowledge of $A$. However, this is of
no worry in the lattice theory, where
one still has to take a continuum limit by letting the
edge length (the lattice spacing) go to zero. In this limit, the usual
canonical commutators are recovered, and -- if all goes well -- 
also commutators of more complicated composite quantities [15].
The same non-vanishing commutators are encountered in the continuum theory, 
where however no further continuum limit is to be taken, because one never 
started from an {\it approximation} to the theory, and there was nothing in the 
course of the construction suggesting the need for such a limit.

\vskip1cm
\line{\bf 3.4 Some current affairs\hfil}

Maybe the problem described in the last section points toward
some further trouble ahead for the continuum loop
representation. More likely perhaps, it can be fixed by
imitating more closely the lattice construction for operators and/or 
selecting a suitable subspace of physical wave functions.
One important difference between the lattice
and continuum approaches is the fact that the lattice operators are
maximally ``adapted" to the wave functions they act on, since they share
from the outset the same support, provided by the links of the finite
lattice. In other words, the regularization takes care of both states
and operators simultaneously. As a consequence, it avoids some 
problematic features of the continuum approach. 

Firstly, on the lattice only a few operators can be defined
which act truly locally, in the sense that they only change the flux-line
routings individually at intersections, without changing the flux-line or spin
assignments. This tends to be the case for operators that are written
purely as functions of the integrated link momenta, and not of the holonomies,
but clearly not for the discretized Hamiltonian operator, say. 
Therefore the dynamics automatically introduces a coupling between
neighbouring 
lattice intersections, which was to be expected on general grounds to
render the theory non-trivial (the same is true for Hamiltonian lattice
gauge theory). The fact that such a coupling is not present for the
Hamiltonians constructed so far in the continuum (including 
Thiemann's otherwise very interesting proposal [20]), seems worrying, 
as has also been remarked elsewhere [21]. 
By the same token, the coupling effect present in the lattice theory
obviously complicates calculations, but this may be unavoidable to
obtain non-trivial results.

Another difficulty one meets in the continuum theory is
the fact that -- although one can define a length operator -- its
interpretation and 
spectral properties are incompatible with those of the area and
volume operators [19] (in fact, a somewhat milder incompatibility is
already present in the case of area and volume alone, as far as their
zero-spectrum is concerned). No such problems occur in the lattice
approach, and first calculations suggest that qualitatively the
spectra of all geometric operators are perfectly 
compatible [16].

\vskip2cm
\line{\ch 4 Outlook\hfil}

I have given you a rather sketchy and biased summary of some
developments taking place in the field of non-perturbative quantum
gravity, and involving discrete methods. The 
existing quantization approaches are rather different, and it is a
challenge to come up with questions that could be answered and compared
in all of them. In the absence of any experimentally
verifiable results, this kind of consistency check seems crucial. 

There are genuinely new results and techniques both in the Regge calculus
schemes and the canonical connection formulation. 
A very interesting point of comparison is the way in which 
diffeomorphism-invariance is handled, 
and in this regard the recent mathematical
construction backing up the dynamical triangulations approach is
most intriguing [11]. 
The connection lattice approach is still incomplete,
and nothing is as yet known about the nature of the spectrum of
the Hamiltonian constraint and the continuum limit. 
New candidates for solutions to the Wheeler-DeWitt equation have been uncovered 
in the continuum [20], but these may yet again turn out 
to be too simple, 
for the reasons outlined in subsection 3.4.

On the whole, given the slow speed with which the subject has advanced over the
last decades, I think the outlook right now is reasonably optimistic: 
various technical frameworks are around that have not yet met any apparent
insurmountable obstacles. Interesting physical questions, for example, on
the nature of the continuum limit and observables, are currently 
being addressed. There is still room for hope in quantum gravity!

\vskip2cm
\line{\ch References\hfil}

\item{[1]} J.F. Doneghue: General relativity as an effective 
 field theory: the leading quantum corrections, 
 \Journal{\PRD}{50}{3874-88}{1994}.

\item{[2]} P. Menotti and A. Pelissetto: Poincar\'e, de Sitter, and 
 conformal gravity on the lattice, \Journal{\PRD}{35}{1194-204}{1987};
 R. Percacci: The Higgs phenomenon in quantum gravity, 
 \Journal{\NPB}{353}{271-90}{1991}.

\item{[3]} A. Ashtekar: New variables for classical and quantum
  gravity, \Journal{\PRL}{57}{2244-7}{1986}; A new 
  Hamiltonian formulation of general relativity, 
  \Journal{\PRD}{36}{1587-1603}{1987}.

\item{[4]} J.F. Barbero G.: Real Ashtekar variables for Lorentzian
  signature space-times, \Journal{\PRD}{51}{5507-10}{1995}.

\item{[5]} P. Renteln and L. Smolin: A lattice approach to spinorial
  quantum gravity, \Journal{\CQG}{6}{275-94}{1989};
  P. Renteln: Some results of SU(2) spinorial lattice
  gravity, \Journal{\CQG}{7}{493-502}{1990};
  O. Bostr\"om, M. Miller and L. Smolin: A new discretization of
  classical and quantum general relativity, Syracuse U. {\it preprint}
  SU-GP-93-4-1;
  R. Loll: Non-perturbative solutions for lattice quantum gravity,
  \Journal{\NPB}{444}{619-39}{1995};
  K. Ezawa: Multi-plaquette solutions for discretized
  Ashtekar gravity, \Journal{\MPLA}{11}{2921-32}{1996};
  H. Fort, R. Gambini and J. Pullin: Lattice knot theory and quantum
  gravity in the loop representation, Penn State U. {\it preprint}
  CGPG-96/8-1.

\item{[6]} T. Regge: General Relativity without coordinates, 
\Journal{\NCA}{19}{558-71}{1961}.

\item{[7]} P.A. Tuckey and R.M. Williams: Regge calculus: a brief
 review and bibliography, \Journal{\CQG}{9}{1409-22}{1992}.

\item{[8]} H.W. Hamber and R.M. Williams: Gauge invariance in 
 simplicial gravity, Cambridge U. {\it preprint} DAMTP-96-68.
  
\item{[9]} H.W. Hamber: Phases of simplicial quantum gravity in 
 four dimensions: estimates for the critical exponents, 
 \Journal{\NPB}{400}{347-89}{1993}.

\item{[10]} B. Br\"ugmann and E. Marinari: Monte Carlo simulations 
 of 4d simplicial quantum gravity, \Journal{\JMP}{36}{6340-52}{1995}; 
 J. Ambj\o rn and J. Jurkiewicz: Scaling in 
 four dimensional quantum gravity, \Journal{\NPB}{451}{643-76}{1995}.

\item{[11]} J. Ambj\o rn, M. Carfora and A. Marzuoli: The 
 geometry of dynamical triangulations, Pavia U. {\it preprint}
 DFNT-PAVIA-9-96.

\item{[12]} M.P. Reisenberger: A left-handed simplicial action for 
 Euclidean general relativity, Vienna {\it preprint} ESI-373, 1996.

\item{[13]} R. Loll: The volume operator in discretized quantum
  gravity, \Journal{\PRL}{75}{3048-51}{1995}.
  
\item{[14]} R. Loll: Spectrum of the volume operator in
  quantum gravity, \Journal{\NPB}{460}{143-54}{1996}.

\item{[15]} R. Loll: Further results on geometric operators in 
 quantum gravity, Potsdam {\it preprint} AEI-023, 1996.

\item{[16]} R. Loll: Latticing quantum gravity, Potsdam {\it preprint}
 AEI-024, 1997.

\item{[17]} R. Loll: A real alternative to quantum gravity in loop
  space, {\it Phys. Rev.} D54 (1996) 5381-4.

\item{[18]} C. Rovelli and L. Smolin: Discreteness of area and 
  volume in quantum gravity, \Journal{\NPB}{442}{593-622}{1995};
 Err. \Journal{\it ibid.}{456}{753-4}{1995}.

\item{[19]} T. Thiemann: A length operator for canonical quantum 
  gravity, Harvard U. {\it preprint} HUTMP-96/B-354.

\item{[20]} T. Thiemann: Anomaly-free formulation of nonperturbative 
  four-dimensional Lorentzian quantum gravity, 
  \Journal{\PLB}{380}{257-64}{1996}; 
  Quantum spin dynamics I\&II, Harvard U. 
  {\it preprints} HUTMP-96/B-351 and B-352.

\item{[21]} L. Smolin: The classical limit and the form of the 
  hamiltonian constraint in nonperturbative quantum gravity,
  Penn State U. {\it preprint} CGPG-96/9-4.

\item{[22]} E. Witten: 2+1 dimensional gravity as an exactly 
 soluble system, \Journal{\NPB}{311}{46-78}{1989}.

\item{[23]} A. Ashtekar {\it et al.}: 2+1 gravity as a toy model 
 for the 3+1 theory, \Journal{\CQG}{6}{L185-93}{1989}.

\item{[24]} J.F. Barbero: Solving the constraints of general 
 relativity, \Journal{\CQG}{12}{L5-10}{1995};
 Reality conditions and Ashtekar variables: a different perspective,
 \Journal{\PRD}{51}{5498-506}{1995}.

\item{[25]} S. Holst: Barbero's Hamiltonian derived from a 
 generalized Hilbert-Palatini action, \Journal{\PRD}{53}{5966-9}{1996}.

\item{[26]} C. Rovelli and L. Smolin: Loop space representation of
  quantum general relativity, \Journal{\NPB}{331}{80-152}{1990}.

\item{[27]} A. Ashtekar and C.J. Isham: Representations of the holonomy
 algebras of gravity and non-Abelian gauge theories, 
 \Journal{\CQG}{9}{1433-67}{1992}.

\item{[28]} A. Ashtekar and J. Lewandowski: Differential geometry on the 
 space of connections modulo gauge transformations, 
 \Journal{\JGP}{17}{191-230}{1995};
 D. Marolf and J.M. Mour\~ao: On the support of the Ashtekar-Lewandowski
 measure, \Journal{\CMP}{170}{583-605}{1995};
 A. Ashtekar {\it et al.}: Quantization of diffeomorphism invariant 
 theories, \Journal{\JMP}{36}{6456-93}{1995}; further
 references are given in these papers.

\item{[29]} J.C. Baez: Link invariants, holonomy algebras and functional
 integration, \Journal{\JFA}{127}{108-31}{1995}.

\item{[30]} C. Rovelli and L. Smolin: Spin networks and quantum 
  gravity, \Journal{\PRD}{52}{5743-59}{1995};
  J.C. Baez: Spin network states in gauge theory, 
  \Journal{\AM}{117}{253}{1996};
  Spin networks and nonperturbative quantum gravity, 
  UC Riverside {\it preprint}, 1995.

\end